\newtheorem{definition}{Definition}

\newtheorem{proposition}{Proposition}
\newtheorem{remark}{Remark}

\newenvironment{proof}[1][Proof]{\noindent\textbf{#1.} }{\ \rule{0.5em}{0.5em}}

\documentclass[12pt]{iopart}

\usepackage{amsfonts}
\usepackage{amssymb}
\begin{document}

\title[Nonlocal aspects of $\lambda$-symmetries]{Nonlocal aspects of $\lambda$-symmetries and ODEs reduction}

\author{D. Catalano Ferraioli}

\address{Dipartimento di Matematica, Universit\`{a} di Milano, via Saldini 50, I-20133 Milano, Italy}

\ead{catalano@mat.unimi.it }
\begin{abstract}

A reduction method of ODEs not possessing Lie point symmetries makes use of
the so called $\lambda$-symmetries (C. Muriel and J. L. Romero, \emph{IMA J.
Appl. Math.} \textbf{66}, 111-125, 2001). The notion of covering for an ODE
$\mathcal{Y}$ is used here to recover $\lambda$-symmetries of $\mathcal{Y}$ as
nonlocal symmetries. In this framework, by embedding $\mathcal{Y}$ into a
suitable system $\mathcal{Y}^{\prime}$ determined by the function $\lambda$,
any $\lambda$-symmetry of $\mathcal{Y}$ can be recovered by a local symmetry
of $\mathcal{Y}^{\prime}$. As a consequence, the reduction method of Muriel
and Romero follows from the standard method of reduction by differential
invariants applied to $\mathcal{Y}^{\prime}$.

\end{abstract}



\section{Introduction}

Local symmetries play an important role in the study of differential
equations. In particular, they are extensively used in the case of ordinary
differential equations (ODEs) since they provide a unified approach to the
reduction problem. In fact, if an ODE $\mathcal{Y}$ admits a local symmetry,
one can use this symmetry to reduce the order of $\mathcal{Y}$ by one. Hence,
when the symmetry algebra of $\mathcal{Y}$ is sufficiently large (and
solvable), this can be solved by quadratures.

Local (classical or higher) symmetries of a $k$-th order ODE $\mathcal{Y}$ in
the unknown $u$ are defined by the solutions to a linear PDEs depending on the
derivatives of $u$ up to order $k-1$. Since the general solution to this PDE
cannot be found unless one knows the general solution to $\mathcal{Y}$, one
usually can only search for particular solutions depending on derivatives of
$u$ up to order $k-2$. Therefore, if $\mathcal{Y}$ does not have such a kind
of symmetries it may be not solvable by quadratures. Moreover, one may
encounter equations which can be solved by quadrature but with a lack of
symmetries of order less or equal to $k-2$. In fact, examples of this type are
well known in the recent literature
(\cite{AbrahamShrauner-Govinder-Leach,Bluman-Kumei,Bluman-Reid,Gonzalez-Gonzalez,Govinder-Leach,Ibragimov-russo,Muriel-Romero-01,Muriel-Romero-03}%
) but the first examples probably go back to equations of the type classified
by Painlev\'{e} \cite{Painleve,Ince}. These examples seem to prove that local
symmetries are sometimes inadequate to handle equations which have not enough
local symmetries and they raise the question of whether an extension of the
notion of symmetry would lead to a more effective method of reduction. Hence,
various attempts in this direction have been done and some new classes of
symmetries have been introduced. Among these, in the last few years, a special
attention has been devoted to a new class of symmetries introduced by Muriel
and Romero in \cite{Muriel-Romero-01} (see also
\cite{Cicogna-Gaeta-Morando,Gaeta-Morando-2,Muriel-Olver,Pucci-Saccomandi}%
).These symmetries are neither Lie point nor Lie-B\"{a}cklund and are called
$\lambda$-symmetries, since they are vector fields which depend on a function
$\lambda$.

In the case of equations with a lack of point symmetries, Muriel and Romero
have shown that many of the order-reduction processes can be explained by the
invariance of the equation under $\lambda$-symmetries. In fact, if an equation
is invariant under a $\lambda$-symmetry, one can obtain a complete set of
functionally independent invariants and reduce the order of the equation by
one as for Lie symmetries.

The aim of this paper is to show that this new class of symmetries can be
always recovered by a class of nonlocal symmetries of the given equation. In
fact, we show that by embedding a given ODE $\mathcal{Y}$ into a suitable
system $\mathcal{Y}^{\prime}$ determined by the function $\lambda$, any
$\lambda$-symmetry of $\mathcal{Y}$ correspond to a local standard symmetry of
$\mathcal{Y}^{\prime}$ (see Proposition \ref{Main_proposition}). As a
consequence, we show that the reduction method of Muriel and Romero follows
from the standard method of reduction by differential invariants applied to
$\mathcal{Y}^{\prime}$.

An outline of the paper is as follows. In section \ref{preliminari}, in order
to fix notations and for the convenience of the reader, we collect some
notations and basic facts from the geometric theory of differential equations.
In section \ref{Interpretazione}, after recalling the definition of $\lambda
$-symmetries (as given in \cite{Muriel-Romero-01}), we present our main result
(Proposition \ref{Main_proposition}) on the nonlocal interpretation of
$\lambda$-symmetries. Finally, in section \ref{riduzione}, we use the nonlocal
interpretation of $\lambda$-symmetries to reinterpret the Muriel-Romero
reduction method as a nonlocal symmetry-reduction method.

\section{Preliminaries on local symmetries\label{preliminari}}

In this section we collect some notations and basic facts from the geometric
theory of differential equations. The reader is referred to
\cite{AVL,Bluman-Anco,Bluman-Kumei,Olver1,Olver2, Vin-et-al,Vino-89} for
further details.

Let $M$ be a smooth manifold and $\pi:E\rightarrow M$ a smooth $r$ dimensional
vector bundle. We denote by $\pi_{k}:J^{k}(\pi)\rightarrow M$ the $k$-order
\emph{jet bundle} associated to $\pi$ and by $j_{k}(s)$ the $k$-order
\emph{jet prolongation} of a section $s$ of $\pi$. Since in this paper we are
only concerned with the case $\dim M=1$, we assume that $M$ and $E$ have local
coordinates $t$ and $(t,u^{1},...,u^{r})$, respectively. Correspondingly, the
induced natural coordinates on $J^{k}(\pi)$ will be $(t,u_{i}^{a})$, $1\leq
a\leq r$, $i=0,1,...,k$, where the $u_{i}^{a}$'s are defined by $u_{i}%
^{a}(j_{\infty}(s))=d^{i}\left(  u^{a}(s)\right)  /dt^{i}$, for any section
$s$ of $\pi$. Moreover, when no confusion arises, Einstein summation
convention over repeated indices will be used.

The $k$\emph{-order jet space} $J^{k}(\pi)$ is a manifold equipped with the
smooth distribution $\mathcal{C}^{k}$ of tangent planes to graphs of $k$-order
jet prolongations $j_{k}(s)$. This is the \emph{contact (or Cartan)
distribution} \emph{of} $J^{k}\left(  \pi\right)  $, whose infinitesimal
symmetries will be referred to as \emph{Lie symmetries} on $J^{k}(\pi)$. A
special kind of Lie symmetries on $J^{k}(\pi)$ is represented by the \emph{Lie
point symmetries} which are obtained as prolongation of vector fields $X$ on
$E$. A fundamental theorem due to B\"{a}cklund \cite{Backlund,Ibragimov-trad}
shows that only when $r=1$ there are examples of Lie symmetries on $J^{k}%
(\pi)$ which are not the prolongation of point transformations. These Lie
symmetries, which do not come from point transformations, are called \emph{Lie
contact symmetries} and can always be recovered as the prolongation of a Lie
symmetry $X$ on $J^{1}(\pi)$. We denote by $X^{(k)}$ (or $X^{(k-1)}$) the Lie
point (or contact, respectively) symmetry obtained by prolonging the vector
field $X$ to $J^{k}(\pi)$.

In this framework a $k$-th order system of differential equations can be
regarded as a submanifold $\mathcal{E}\subset J^{k}(\pi)$ and any solution of
the system is a section of $\pi$ whose $k$-order prolongation is an integral
manifold of the restriction $\left.  \mathcal{C}^{k}\right\vert _{\mathcal{E}%
}$ of the contact distribution to $\mathcal{E}$. A Lie symmetry which is
tangent to $\mathcal{E}$ is, of course, also a symmetry of $\left.
\mathcal{C}^{k}\right\vert _{\mathcal{E}}$ and is called a \emph{classical
symmetry of} $\mathcal{E}$. The fundamental role of classical symmetries in
this context is due to the fact that, since they shuffle integral manifolds of
$\left.  \mathcal{C}^{k}\right\vert _{\mathcal{E}}$, they induce a
transformation on the space of solutions of $\mathcal{E}$.

The natural projections $\pi_{h,k}:J^{h}(\pi)\rightarrow J^{k}(\pi)$, for any
$h>k$, allow one to define the bundle of infinite jets $J^{\infty}%
(\pi)\rightarrow M$ as the inverse limit of the tower of projections
$M\longleftarrow E\longleftarrow J^{1}(\pi)\longleftarrow J^{2}(\pi
)\longleftarrow...$.

The manifold $J^{\infty}(\pi)$ is infinite dimensional with induced
coordinates $(t,u_{i}^{a})$, $1\leq a\leq r$, $i=0,1,...$, and the
$\mathbb{R}$-algebra of smooth functions on $J^{\infty}(\pi)$ is defined as
$\mathcal{F}(\pi)=\cup_{l}C^{\infty}(J^{l}(\pi))$. Hence, any $f\in
\mathcal{F}(\pi)$ is a function of some arbitrary large but finite number of
jet coordinates. Analogously to the case of finite dimensional manifolds,
tangent vectors and vector fields on $J^{\infty}(\pi)$ are defined as
derivations of $\mathcal{F}(\pi)$. The set $D(\pi)$ of vector fields on
$J^{\infty}(\pi)$ has the structure of a Lie algebra, with respect to the
commutator $\left[  \rule{4pt}{0pt},\rule{4pt}{0pt}\right]  $.

Since smooth sections of $\pi$ can be infinitely prolonged, if we consider the
tangent planes to the graphs of $\infty$-order jet prolongations $j_{\infty
}(s)$, one can also define a \emph{contact distribution }$\mathcal{C}$ on
$J^{\infty}(\pi)$. In terms of coordinates, $\mathcal{C}$ is defined by the
\emph{total derivative} operator
\[
D=\partial_{t}+u_{i+1}^{a}\partial_{u_{i}^{a}}%
\]
which annihilates all the \emph{contact forms} $\theta_{s}^{a}=du_{s}%
^{a}-u_{s+1}^{a}dt$.

If $X$ is a Lie symmetry, by considering the sequence of prolongations
$X^{(1)},X^{(2)},...$ one gets the vector field $X^{(\infty)}$, which is a
symmetry of $\mathcal{C}$ called the \emph{infinite prolongation} of $X$.
However, contrary to the case of finite order jet spaces, symmetries of
$\mathcal{C}$ cannot always be recovered by infinite prolongations of Lie
symmetries. In fact it can be proved that $Y=\xi\partial_{t}+\eta_{i}%
^{a}\partial_{u_{i}^{a}}$ is an infinitesimal symmetry of $\mathcal{C}$ if and
only if $\xi,\eta_{0}^{a}\in\mathcal{F}(\pi)$ are arbitrary functions and
\begin{equation}
\eta_{i}^{a}=D(\eta_{i-1}^{a})-D(\xi)u_{i}^{a}. \label{Y_symm}%
\end{equation}
Hence, $Y$ is the infinite prolongation of a Lie point (or contact) symmetry
iff $\xi,\eta_{0}^{a}$ are functions on $E$ (or $J^{1}(\pi)$, respectively).

Given a differential equation $\mathcal{E}\subset J^{k}(\pi)$, the
$l$\emph{-th prolongation of }$\mathcal{E}$ is the set of points
$\mathcal{E}^{(l)}:=\left\{  j_{k+l}(s)(x)\right\}  $ $\subset$ $J^{k+l}(\pi)$
such that the Taylor expansion up to order $k+l$, at the point $x\in M$, of
the section $s$ satisfies the equation $\mathcal{E}$. Analogously, by
considering the infinite prolongations of sections of $\pi$, one can define
the \emph{infinite} \emph{prolongation} $\mathcal{E}^{\infty}$.

In this paper we will deal only with (systems of) ordinary differential
equations $\mathcal{E}$ which are in \emph{normal form\footnote{That is,
$\mathcal{E}$ is solved with respect to the higher derivatives.}} and
\emph{not underdetermined}. Hence, if $\mathcal{E}$ is defined by some smooth
(vector) function $F=0$, $\mathcal{E}^{\infty}$ is finite dimensional and
defined by the infinite system of equations%
\[
D^{s}(F)=0,\qquad s=0,1,....
\]
In this case, symmetries $Y$ of $\mathcal{C}$ which are tangent to
$\mathcal{E}^{\infty}$ are called \emph{higher symmetries} of $\mathcal{E}$
and are determined by the condition $\left.  Y(F)\right\vert _{\mathcal{E}%
^{\infty}}=0$, i.e. that $Y(F)$ vanishes when restricted to $\mathcal{E}%
^{\infty}$. In particular, any higher symmetry of $\mathcal{E}$ is an
infinitesimal symmetry of the restriction $\left.  \mathcal{C}\right\vert
_{\mathcal{E}^{\infty}}$ of the contact distribution to $\mathcal{E}^{\infty}$.

Analogously to the case of classical symmetries, the key role played by
symmetries of $\mathcal{E}^{\infty}$ is due to the fact that they shuffle
integral manifolds of $\left.  \mathcal{C}\right\vert _{\mathcal{E}^{\infty}}$
and hence they induce a transformation in the space of infinite prolongations
of solutions to $\mathcal{E}$. Hence, instead of computing symmetries $Y$ of
$\mathcal{C}$ which are tangent to $\mathcal{E}^{\infty}$, one should be
mainly interested in computing their restrictions $\left.  Y\right\vert
_{\mathcal{E}^{\infty}}$ to $\mathcal{E}^{\infty}$. This choice turns out to
be also convenient since it noteworthy simplifies computations (see
\cite{Vin-et-al,Vino-89} for more details about these and others aspects of
$\infty$-jets theory).

\section{Nonlocal interpretation of $\lambda$%
-symmetries\label{Interpretazione}}

This section is devoted to the proof of the main result of the paper, which is
the reinterpretation of $\lambda$-symmetries of an ODE $\mathcal{Y}$ (as
introduced in \cite{Muriel-Romero-01}) as shadows of nonlocal symmetries of
$\mathcal{Y}$.

\subsection{Nonlocal symmetries\label{sez_symm_nonloc}}

A first heuristic generalization of local symmetries appeared in
\cite{Konopelchenko} and \cite{Kapcov} in the form of a nonlocal point of
view. A conceptual framework for nonlocal symmetries is provided by the notion
of \emph{covering}. Since here we only deal with ordinary differential
equations, we give the definition of covering only in this case. The
interested reader, is referred to \cite{Kras-Vin,Vin-et-al} for the general
definition and further details.

\begin{definition}
Let $\mathcal{Y}$ be a $k$-order ODE on a one-dimensional bundle $\pi_{0}$. We
shall say that a smooth bundle $\kappa:\mathcal{\widetilde{Y}\rightarrow
Y}^{\infty}$ is a covering for the ordinary differential equation
$\mathcal{Y}$ if the manifold $\mathcal{\widetilde{Y}}$ is equipped with a
1-dimensional distribution%
\[
\mathcal{\widetilde{C}=}\left\{  \mathcal{\widetilde{C}}_{p}\right\}
_{p\in\mathcal{\widetilde{Y}}}%
\]
and, for any point $p\in\mathcal{\widetilde{Y}}$, the tangent mapping
$\kappa_{\ast}$ gives an isomorphism between $\mathcal{\widetilde{C}}$ and the
restriction $\left.  \mathcal{C}\right\vert _{\mathcal{Y}^{\infty}}$ of the
contact distribution of $J^{\infty}(\pi_{0})$ to $\mathcal{Y}^{\infty}$.
\end{definition}

The dimension of the bundle $\kappa$ is called the dimension of the covering
and is denoted by $\dim(\kappa)$. Below we are mainly concerned with finite
dimensional coverings and, in particular, with the case $\dim(\kappa)=1$.

Given a covering $\kappa$, integral manifolds of $\mathcal{\widetilde{C}}$
play a special role in the geometry of $\mathcal{Y}$. In fact, any integral
manifold $\widetilde{\Sigma}$ of $\mathcal{\widetilde{C}}$ projects, through
$\kappa$, to an integral manifold of $\left.  \mathcal{C}\right\vert
_{\mathcal{Y}^{\infty}}$, i.e., to a (possible degenerating) solution of the
equation $\mathcal{Y}$. However, this correspondence between solutions of
$\mathcal{Y}$ and integral manifolds of $\mathcal{\widetilde{C}}$ is not
one-to-one, since, if $\Sigma$ is an integral manifold of $\left.
\mathcal{C}\right\vert _{\mathcal{Y}^{\infty}}$, then $\kappa^{-1}(\Sigma)$ is
a family of integral manifolds of $\mathcal{\widetilde{C}}$. In particular,
one can interpret $\kappa^{-1}(\Sigma)$ as a parameterized family of solutions
to $\mathcal{Y}$. It follows that, in this picture, symmetries of
$\mathcal{\widetilde{C}}$ also play a key role. In fact, since these
symmetries shuffle integral manifolds of $\mathcal{\widetilde{C}}$, they
induce a transformation on $\mathcal{Y}$ which maps solutions to solutions. In
view of this fact, we give the following

\begin{definition}
Nonlocal symmetries of $\mathcal{Y}$ are the symmetries of the distribution
$\mathcal{\widetilde{C}}$ of a covering $\kappa:\mathcal{\widetilde
{Y}\rightarrow Y}^{\infty}$.
\end{definition}

Let $\pi_{0}$ be the trivial one-dimensional bundle over $\mathbb{R}$, with
standard coordinates $(t,v)$, and $\mathcal{Y}$ be given as
\begin{equation}
v_{k}=f(t,v,v_{1},...,v_{k-1}),\label{equazione}%
\end{equation}
for some smooth function $f$. Below we will consider only coverings where
$\kappa$ is a trivial bundle $\kappa:\mathcal{Y}^{\infty}\times W\rightarrow
\mathcal{Y}^{\infty}$, with $W\subset\mathbb{R}$ ($\dim(\kappa)=1$). In this
case, if $w$ is the standard coordinate in $W$, the distribution
$\mathcal{\widetilde{C}}$ is generated by the vector field on
$\mathcal{\widetilde{Y}}$ defined as
\begin{equation}
\left.  \widetilde{D}\right\vert _{\mathcal{\widetilde{Y}}}=\bar{D}%
_{0}+H\partial_{w}\label{D_widetilde}%
\end{equation}
where $H$ is a smooth function on $\mathcal{Y}^{\infty}\times W$ and $\bar
{D}_{0}=\partial_{t}+v_{1}\partial_{v}+...+f\partial_{v_{k-1}}$ is the
restriction to $\mathcal{Y}^{\infty}$ of the total derivative operator on
$J^{\infty}(\pi_{0})$, i.e.,
\[
D_{0}=\partial_{t}+v_{i+1}\partial_{v_{i}}.
\]
Hence, the covering $\kappa$ is determined by the system $\mathcal{Y}^{\prime
}$ defined by (\ref{equazione}) together with the additional equation
\[
\frac{dw}{dt}=H.
\]
In fact, if one considers the trivial bundle $\widetilde{\pi}:\mathbb{R}%
^{3}\rightarrow\mathbb{R}$ with coordinates $(t,u^{1}=v,u^{2}=w)$, then
$\mathcal{Y}^{\prime}\subset J^{k}(\widetilde{\pi}),\mathcal{\widetilde{Y}%
}=\left(  \mathcal{Y}^{\prime}\right)  ^{\infty}$ and $\kappa$ is the obvious
projection $\left(  \mathcal{Y}^{\prime}\right)  ^{\infty}\rightarrow
\mathcal{Y}^{\infty}$. Moreover, the vector field (\ref{D_widetilde}) is just
the restriction to $\left(  \mathcal{Y}^{\prime}\right)  ^{\infty}$ of the
total derivative operator on $J^{\infty}(\widetilde{\pi})$, i.e.,
\[
\widetilde{D}=D_{0}+w_{i+1}\partial_{w_{i}}.
\]
Nonlocal symmetries of $\mathcal{Y}$ are symmetries of the vector field
(\ref{D_widetilde}) and can be determined through a symmetry analysis of the
system $\mathcal{Y}^{\prime}$. Therefore nonlocal symmetries of $\mathcal{Y}$
are infinitesimal symmetries of the contact distribution on $J^{\infty
}(\widetilde{\pi})$ which are tangent to $\left(  \mathcal{Y}^{\prime}\right)
^{\infty}$ and, in view of (\ref{Y_symm}) (where $D$ now stands for
$\widetilde{D}$), have the form%
\begin{equation}
Y=\xi\partial_{t}+\eta_{i}^{1}\partial_{v_{i}}+\eta_{i}^{2}\partial_{w_{i}%
}\label{pippo}%
\end{equation}
with
\begin{equation}%
\begin{array}
[c]{l}%
\eta_{i}^{1}=\widetilde{D}(\eta_{i-1}^{1})-\widetilde{D}(\xi)v_{i},\\
\eta_{i}^{2}=\widetilde{D}(\eta_{i-1}^{2})-\widetilde{D}(\xi)w_{i}.
\end{array}
\label{formule_prolungamanto_non_locale}%
\end{equation}
In the rest of the paper we will only consider nonlocal symmetries of
$\mathcal{Y}$ with
\begin{equation}
\xi=\xi(t,v,w),\qquad\eta_{0}^{1}=\eta_{0}^{1}(t,v,w),\qquad\eta_{0}^{2}%
=\eta_{0}^{2}(t,v,v_{1},...,v_{k-1},w).\label{pippo_2}%
\end{equation}
Since these symmetries are possibly nonclassical symmetries of $\mathcal{Y}%
^{\prime}$, they are completely determined by a symmetry analysis of $\left(
\mathcal{Y}^{\prime}\right)  ^{\infty}$ on $J^{\infty}(\widetilde{\pi})$ and
will be called \emph{semi-classical nonlocal symmetries} of $\mathcal{Y}$.

\subsection{$\lambda$-Symmetries}

In this subsection we recall the definition of $\lambda$-prolongation and
$\lambda$-symmetries for an ODE $\mathcal{Y}$ as given by Muriel and Romero in
\cite{Muriel-Romero-01}. Here by $\pi_{0}$ we still denote the trivial
one-dimensional bundle over $\mathbb{R}$ with standard coordinates $(t,v)$ and
by $\mathcal{Y}$ a $k$-order ODE of the form (\ref{equazione}) on $\pi_{0}$.

\begin{definition}
[Muriel-Romero]Let $\lambda$ be a smooth function on $J^{1}(\pi_{0})$, then
the $\lambda$-prolongation to $J^{k}(\pi_{0})$ of a vector field
$X=\rho\partial_{t}+\psi\partial_{v}$ on $J^{0}(\pi_{0})$ is the vector field%
\[
X^{[\lambda,k]}=\rho\partial_{t}+\sum_{i=0}^{k}\psi^{\lbrack\lambda
,i]}\partial_{v_{i}}%
\]
with
\[
\psi^{\lbrack\lambda,0]}=\psi,\qquad\psi^{\lbrack\lambda,i]}=D_{0}%
(\psi^{\lbrack\lambda,i-1]})-D_{0}(\rho)v_{i}+\lambda\left(  \psi
^{\lbrack\lambda,i-1]}-\rho v_{i}\right)  ,
\]
and $D_{0}=\partial_{t}+v_{1}\partial_{v}+...+v_{k}\partial_{v_{k-1}}$ the
(truncated) total derivative operator on $J^{k}(\pi_{0})$.
\end{definition}

\begin{remark}
By a straightforward computation \cite{Muriel-Romero-01}, it can be shown that
a vector field $U$ on $J^{k}(\pi_{0})$ is a vector field of the form
$X^{[\lambda,k]}$ if and only if there exist two functions $\mu,\lambda$ on
$J^{1}(\pi_{0})$ such that%
\begin{equation}
\left[  U,D_{0}\right]  =\mu D_{0}+\lambda U. \label{comm_con_lambda}%
\end{equation}

\end{remark}

\begin{definition}
[Muriel-Romero]We say that a vector field $X^{[\lambda,k]}$, for some function
$\lambda$, is a $\lambda$-symmetry of $\mathcal{Y}$ if and only if
$X^{[\lambda,k]}$ is tangent to $\mathcal{Y}$.
\end{definition}

\subsection{Nonlocal interpretation of $\lambda$-symmetries}

In this subsection we will provide a characterization of $\lambda$-symmetries
of $\mathcal{Y}$ in terms of semi-classical nonlocal symmetries, i.e.,
nonlocal symmetries of the form (\ref{pippo})-(\ref{pippo_2}). In fact, using
the same notations of subsection \ref{sez_symm_nonloc} above, by considering
the covering $\kappa_{\lambda}:\left(  \mathcal{Y}^{\prime}\right)  ^{\infty
}\rightarrow\mathcal{Y}^{\infty}$ defined by the ODE system%
\begin{equation}
\mathcal{Y}^{\prime}:=\left\{  v_{k}=f,w_{1}=\lambda\right\}  \label{Y'}%
\end{equation}
one gets the following

\begin{proposition}
\label{Main_proposition}An ODE $\mathcal{Y}$ admits a $\lambda$-symmetry
$X^{[\lambda,k]}$ if and only if $\mathcal{Y}$ admits a semi-classical
nonlocal symmetry $Y$ which is a symmetry of (\ref{Y'}) such that $\left[
\partial_{w},Y\right]  =Y$.
\end{proposition}

\begin{proof}
Let $X^{[\lambda,k]}$ be a $\lambda$-symmetry of $\mathcal{Y}$ which is the
$\lambda$-prolongation to $J^{k}(\pi_{0})$ of a vector field $X=\rho
\partial_{t}+\psi\partial_{v}$ on $J^{0}(\pi_{0})$. We show that one can
determine a function $\chi=\chi(t,v,v_{1},...,v_{k-1})$ such that the vector
field
\begin{equation}
Y=e^{w}\rho\partial_{t}+\eta_{i}^{1}\partial_{v_{i}}+\eta_{i}^{2}%
\partial_{w_{i}}, \label{Y_corresp}%
\end{equation}
with $\eta_{0}^{1}=e^{w}\psi,\eta_{0}^{2}=e^{w}\chi$ and
\begin{equation}%
\begin{array}
[c]{l}%
\eta_{i}^{1}=\widetilde{D}(\eta_{i-1}^{1})-\widetilde{D}(e^{w}\rho)v_{i},\\
\eta_{i}^{2}=\widetilde{D}(\eta_{i-1}^{2})-\widetilde{D}(e^{w}\rho)w_{i},
\end{array}
\label{Y_corresp_bis}%
\end{equation}
is a semi-classical nonlocal symmetry of $\mathcal{Y}$ satisfying $\left[
\partial_{w},Y\right]  =Y$. In fact, one can readily show that
(\ref{Y_corresp}) has the form (\ref{pippo})-(\ref{pippo_2}) and satisfies
condition $\left[  \partial_{w},Y\right]  =Y$. Hence we only need to show
that, for a suitable choice of $\chi$, $Y$ is also tangent to $\left(
\mathcal{Y}^{\prime}\right)  ^{\infty}$. To this end we notice that the
restriction of the function $Y(v_{k}-f)$ to $\left(  \mathcal{Y}^{\prime
}\right)  ^{\infty}$ agrees with the restriction of $X^{[\lambda,k]}(v_{k}-f)$
to $\mathcal{Y}$. Therefore, since , $X^{[\lambda,k]}$ is tangent to
$\mathcal{Y}$, one gets that $\left.  Y(v_{k}-f)\right\vert _{\left(
\mathcal{Y}^{\prime}\right)  ^{\infty}}=0$ and $Y$ is tangent to $\left(
\mathcal{Y}^{\prime}\right)  ^{\infty}$ iff $\left.  Y(w_{1}-\lambda
)\right\vert _{\left(  \mathcal{Y}^{\prime}\right)  ^{\infty}}=\allowbreak
\left.  \left(  \eta_{1}^{2}-e^{w}X^{[\lambda,1]}(\lambda)\right)  \right\vert
_{\left(  \mathcal{Y}^{\prime}\right)  ^{\infty}}=0$. Hence, $Y$ is tangent to
$\left(  \mathcal{Y}^{\prime}\right)  ^{\infty}$ iff $\chi$ is a solution of
the following linear first-order equation on $\left(  \mathcal{Y}^{\prime
}\right)  ^{\infty}$%
\begin{equation}
\partial_{t}\chi+\sum_{i=1}^{k-1}v_{i}\partial_{v_{i-1}}\chi+f\partial
_{v_{k-1}}\chi=X^{[\lambda,1]}(\lambda)+\lambda^{2}\rho+\lambda\left(
\partial_{t}\rho+v_{1}\partial_{v}\rho-\chi\right)  .\qquad\label{eq_*}%
\end{equation}
Since for any non-characteristic Cauchy data (\ref{eq_*}) admits a unique
solution (see \cite{Arnold} or \cite{Smirnov}), one eventually gets that any
$\lambda$-symmetry of $\mathcal{Y}$ determines a semi-classical nonlocal
symmetry $Y$ of $\mathcal{Y}$ such that $\left[  \partial_{w},Y\right]
=Y$.\newline Conversely, let $Y$ be a semi-classical nonlocal symmetry $Y$ of
$\mathcal{Y}$ with respect to the covering $\kappa_{\lambda}$. Since any such
symmetry has the form (\ref{pippo})-(\ref{pippo_2}), condition $\left[
\partial_{w},Y\right]  =Y$ is satisfied if and only if the functions $\xi
,\eta_{i}^{1}$ and $\eta_{i}^{2}$ have the form%
\begin{equation}
\xi=e^{w}\rho,\eta_{i}^{1}=e^{w}\psi_{i},\eta_{i}^{2}=e^{w}\chi_{i}
\label{coeff_exp}%
\end{equation}
with $\rho,\psi_{i}$ functions of $(t,v)$ and $\chi_{i}$ a function of
$(t,v,v_{1},...,v_{k-1})$.\newline Then, plugging (\ref{coeff_exp}) into
(\ref{formule_prolungamanto_non_locale}) and using the definition
$\widetilde{D}=D_{0}+\sum_{i}w_{i+1}\partial_{w_{i}}$(here $D_{0}$ is the
total derivative operator on $J^{\infty}(\pi_{0})$), one gets
\begin{equation}%
\begin{array}
[c]{l}%
\psi_{i}=D_{0}(\psi_{i-1})-D_{0}(\rho)v_{i}+w_{1}\left(  \psi_{i-1}-\rho
v_{i}\right)  ,\\
\chi_{i}=D_{0}(\chi_{i-1})-D_{0}(\rho)w_{i}+w_{1}\left(  \chi_{i-1}-\rho
w_{i}\right)  .
\end{array}
\label{lambda_to_symm}%
\end{equation}
Now, if one puts $X=\rho\partial_{t}+\psi_{0}\partial_{v}$,
(\ref{lambda_to_symm}) entails that the restriction of $X^{[\lambda,k]}\left(
v_{k}-f\right)  $ to $\mathcal{Y}$ agrees with the restriction of
$e^{-w}Y\left(  v_{k}-f\right)  $ to $\left(  \mathcal{Y}^{\prime}\right)
^{\infty}$. Therefore, by tangency of $Y$ to $\left(  \mathcal{Y}^{\prime
}\right)  ^{\infty}$, it follows that $X^{[\lambda,k]}$ is tangent to
$\mathcal{Y}$. This concludes the proof, since it proves that any
semi-classical nonlocal symmetry $Y$ of $\mathcal{Y}$, satisfying condition
$\left[  \partial_{w},Y\right]  =Y$, returns a $\lambda$-symmetry
$X^{[\lambda,k]}$ of $\mathcal{Y}$.
\end{proof}

\begin{remark}
\label{Remark_1}This proposition proves that, by using formulas
(\ref{Y_corresp})-(\ref{Y_corresp_bis}) and (\ref{eq_*}), it is always
possible to reconstruct a nonlocal symmetry $Y$ from any given $\lambda
$-symmetry $X^{[\lambda,k]}$. In this sense, $\lambda$-symmetries can be
geometrically regarded as shadows of nonlocal symmetries. However, we notice
that the problem of determining the general solution of (\ref{eq_*}) (the
reconstruction problem for $X^{[\lambda,k]}$) should be at least as difficult
as solving the given ODE $\mathcal{Y}$. Therefore, in practice it could be not
so easy to determine such a correspondence (see Example 3 below).
Nevertheless, almost all the examples of $\lambda$-symmetries available in
literature can be recovered as nonlocal symmetries with $\chi$ a solution of
(\ref{eq_*}) depending only on $(t,v)$. These examples of $\lambda$-symmetries
are, in fact, shadows of nonlocal symmetries which are classical symmetries of
$\mathcal{Y}^{\prime}$.
\end{remark}

In order to show how in practice $\lambda$-symmetries can be recovered as
nonlocal symmetries, we consider the following examples.\newline%
\textbf{Example 1. }Let $\mathcal{Y}$ be defined as%
\begin{equation}
v_{2}=\frac{v_{1}^{2}}{v}+pg(t)v^{p}v_{1}+g^{\prime}(t)v^{p+1}%
\label{esempio_gonzalez}%
\end{equation}
where $p\neq0$ is a constant. In \cite{Gonzalez}, it has been shown that
(\ref{esempio_gonzalez}) is integrable by quadratures but has Lie point
symmetries only for very restricted forms of the function $g(t)$. On the
contrary, this class of equations admits the $\lambda$-symmetry $X^{[\lambda
,2]}=\partial_{v}+\sum_{i=1}^{2}\psi^{\lbrack\lambda,i]}\partial_{v_{i}}$,
with $\lambda=(pg(t)v^{p+1}+v_{1})/v$, for any form of $g(t)$ (see
\cite{Muriel-Romero-01}). Here we show that the allowed $\lambda$-symmetries
can be recovered as nonlocal symmetries of (\ref{esempio_gonzalez}). To this
end it will suffice to determine a particular solution $\chi=\chi(t,v)$, if
any, of equation (\ref{eq_*}). Now, in the case of equation
(\ref{esempio_gonzalez}), under above assumption on the form of $\chi$,
equation (\ref{eq_*}) reads%
\[
\left(  \partial_{v}\chi+\chi v^{-1}\right)  v_{1}=-\partial_{t}\chi
+v^{-2}\left(  p(p+1)g(t)v^{p+1}-\chi pg(t)v^{p+2}\right)  .
\]
Therefore, $\chi$ is a solution of the following system%
\[
\left\{
\begin{array}
[c]{l}%
\partial_{v}\chi+\chi v^{-1}=0,\rule[-8pt]{0pt}{4pt}\\
\partial_{t}\chi=\left(  p(p+1)g(t)v^{p-1}-\chi pg(t)v^{p}\right)
\end{array}
\right.
\]
and a straightforward computation gives%
\[
\chi=\left(  p+1\right)  v^{-1}.
\]
Then, $\lambda$-symmetries of (\ref{esempio_gonzalez}) can be recovered by
nonlocal symmetries which are the prolongation to $J^{2}(\widetilde{\pi})$ of
the vector field $e^{w}\left(  \partial_{v}+(p+1)/v\partial_{w}\right)
$.\newline\textbf{Example 2. }Let $\mathcal{Y}$ be defined as%
\begin{equation}
v^{5}+e^{2\left(  1/v+t\right)  }\left(  v^{4}+v^{5}-3v_{1}^{2}+vv_{2}\right)
=0.\label{esempio_2}%
\end{equation}
As shown in \cite{Muriel-Romero-03}, (\ref{esempio_2}) has no Lie point
symmetries. Nevertheless, for $\lambda=-v$, (\ref{esempio_2}) has the
$\lambda$-symmetry $X^{[\lambda,2]}$ which is the prolongation of the vector
field $\partial_{t}+v^{2}\partial_{v}$. As above, we show how recover this
$\lambda$-symmetry by a nonlocal symmetry of (\ref{esempio_2}). To this end we
determine a particular solution $\chi=\chi(t,v)$ of equation (\ref{eq_*})
which, in this case, reads%
\[
\partial_{t}\chi-v\chi=-v_{1}\partial_{v}\chi.
\]
Therefore, it is easy to check that in this case $\chi=0$ and the nonlocal
symmetry is the prolongation to $J^{2}(\widetilde{\pi})$ of the vector field
$e^{w}\left(  \partial_{t}+v^{2}\partial_{v}\right)  $.\newline\textbf{Example
3.} Let $\mathcal{Y}$ be defined as
\begin{equation}
v_{2}=\frac{v_{1}^{2}}{v}+\left(  v+\frac{t}{v}\right)  v_{1}%
-1.\label{Esempio 3}%
\end{equation}
This is an instance of a Painlev\`{e}-type equation (see \cite{Ince}) which
does not have Lie point symmetries. Nevertheless, one can readily check that
the $\lambda$-prolonged vector field $X^{[\lambda,2]}$, for $\lambda=v+t/v$
and $X=v\partial_{v}$, is a $\lambda$-symmetry of this equation. In this case,
equation (\ref{eq_*}) reads
\begin{equation}
\partial_{t}\chi+v_{1}\partial_{v}\chi+\left[  \frac{v_{1}^{2}}{v}+\left(
v+\frac{t}{v}\right)  v_{1}-1\right]  \partial_{v_{1}}\chi+\left(  v+\frac
{t}{v}\right)  \chi-v+\frac{t}{v}=0\qquad\label{pippone}%
\end{equation}
and it is easy to check that it does not admit solutions of the form
$\chi=\chi(t,v)$. Therefore, the reconstruction of a nonlocal symmetry $Y$ (of
the form (\ref{Y_corresp})-(\ref{eq_*})) corresponding to $X^{[\lambda,2]}$ is
conditioned to the determination of a solution of (\ref{pippone}) in the
general form $\chi=\chi(t,v,v_{1})$.

\section{Reduction via $\lambda$-symmetries\label{riduzione}}

One of the most important classical application of symmetry analysis to ODEs
is the reduction of order. In particular, one can show that (Lie-Bianchi
Theorem) \cite{Olver1,Vin-et-al} if a $k$-order ODE $\mathcal{Y}$ possesses a
solvable $k$-dimensional Lie algebra of Lie point symmetries, then
$\mathcal{Y}$ is integrable by quadratures. More in general, given a symmetry
(Lie point, contact or higher) of an ODE $\mathcal{Y}$, one can use its
differential invariants to reduce the order of $\mathcal{Y}$
\cite{Bluman-Kumei,Bluman-Anco,Olver1,Olver2}. This procedure, usually
referred to as \emph{the method of differential invariants} (MDI), can be
described as follows. If $\mathcal{Y}\subset J^{k}(\pi_{0})$ is an ODE of the
form (\ref{equazione}), for any symmetry $X$ of $\mathcal{Y}$ the restriction
$\overline{X}$ to $\mathcal{Y}$ (or to $\mathcal{Y}^{\infty}$, if $X$ is a
higher symmetry) has at most $k$ functionally independent differential
invariants, say $\tau,\phi_{0},\phi_{1}...,\phi_{k-2}$. It follows that, since
the restrictions of $X$ and $D_{0}$ (to $\mathcal{Y}$ or $\mathcal{Y}^{\infty
}$) are such that $\left[  \rule{2pt}{0pt}\overline{X},\overline{D_{0}%
}\rule{2pt}{0pt}\right]  =\alpha\overline{D_{0}}$ (for some function $\alpha
$), the restrictions of the functions $g_{i}=D_{0}(\phi_{i})/D_{0}%
(\tau),i=0,1,...,k-2,$ also are differential invariants and must depend on
$\tau,\phi_{0},...,\phi_{k-2}$. Therefore, in terms of these invariants,
$\mathcal{Y}$ can be written as the system of first order $k-1$ equations
$\left\{  d\phi_{i}/d\tau=g_{i}\right\}  $. In particular, if $X$ is a Lie
point symmetry, one can choose the invariants $\tau,\phi_{0},...,\phi_{k-2}$
in such a way that $\tau,\phi_{0}$ are zeroth and first order invariants,
respectively, and $\phi_{i}=D_{0}(\phi_{i-1})/D_{0}(\tau)$. In fact, since
each $\phi_{i}$ depends on $v_{i+1}$, the system of invariants so defined is
independent and complete. Hence $\phi_{k-1}=d\phi_{k-2}/d\tau$ is a function
of $(\tau,\phi_{0},...,\phi_{k-2})$ and $\mathcal{Y}$ takes the form of the
$\left(  k-1\right)  $-order equation $\phi_{k-1}=\phi_{k-1}(\tau,\phi
_{0},...,\phi_{k-2})$.

However, finding symmetries for ODEs is not always easy and often one
encounters equations with a lack of local symmetries. For example, it is well
known that for a $k$-order ODE $\mathcal{Y}$ the general solution to the
determining equation for its $\left(  k-1\right)  $-order symmetries (i.e.,
symmetries on a $(k-1)$-order jet space) cannot be found unless one knows the
general solution to $\mathcal{Y}$. Moreover there are a number of examples
\cite{Bluman-Anco,Bluman-Kumei,Gonzalez-Gonzalez,Govinder-Leach,Ibragimov-russo,Olver2,Muriel-Romero-01,Muriel-Romero-03}
of equations with no symmetries of order less or equal to $k-2$ (in particular
with no Lie point symmetries). Therefore, in these cases the MDI cannot be
implemented and one needs alternative methods. The method proposed by Muriel
and Romero is of this sort since it allows the reduction of ODEs not
possessing Lie point symmetries.

The following Proposition summarizes the method of Muriel and Romero (see
\cite{Muriel-Romero-01} for details).

\begin{proposition}
\label{Rid-Muriel-Romero}Let $X^{[\lambda,k]}$ be a $\lambda$-symmetry of the
equation $\mathcal{Y}=\left\{  \Delta(t,v,...,v_{k})=0\right\}  $, with
$\lambda=\lambda(t,v,v_{1})$, and let $x=x(t,v)$, $\zeta_{0}=\zeta
_{0}(t,v,v_{1})$ be two functionally independent invariants of $X^{[\lambda
,k]}$. The general solution of $\Delta=0$ can be obtained by solving first a
reduced equation of the form $\Delta_{red.}(x,\zeta_{0},...,\zeta_{k-1})=0$,
with $\zeta_{i}=d^{i}\zeta_{0}/dx^{i}$, and then the auxiliary $1$-order ODE
$\zeta_{0}=\zeta_{0}(t,v,v_{1})$.
\end{proposition}

The reduced equation $\Delta_{red.}(x,\zeta_{0},...,\zeta_{k-1})=0$ is
constructed as follows. Since $\zeta_{i}=D_{0}(\zeta_{i-1})/D_{0}(x)$, each
$\zeta_{i}$ depends on $v_{i+1}$ and (in view of (\ref{comm_con_lambda}))
$\left\{  x,\zeta_{0},...,\right.  \allowbreak\left.  \zeta_{k-1}\right\}  $
is a complete set of functionally independent differential invariants of
$X^{[\lambda,k]}$. Therefore, the reduced equation $\Delta_{red.}=0$ is
determined by rewriting $\Delta=0$ in terms of these invariants.

However, since we have proved that any $\lambda$-symmetry of $\mathcal{Y}%
=\{\Delta=0\}$ can be recovered as a local symmetry of $\mathcal{Y}^{\prime
}=\left\{  \Delta=0,w_{1}=\lambda\right\}  $, one can expect that Proposition
\ref{Rid-Muriel-Romero} above represents an application of MDI to
$\mathcal{Y}^{\prime}$. It is indeed the case as the following argument shows.

Under assumptions of Proposition \ref{Rid-Muriel-Romero}, the $\lambda
$-symmetry $X^{[\lambda,k]}$ can be recovered by a nonlocal symmetry $Y$ whose
restriction $\overline{Y}$ to $\left(  \mathcal{Y}^{\prime}\right)  ^{\infty}$
possess the invariants $x=x(t,v),\zeta_{0}=\zeta_{0}(t,v,v_{1})$. Then
$\overline{Y}$ also admits the derived invariants $\zeta_{i}=\overline{D_{0}%
}(\zeta_{i-1})/\overline{D_{0}}(x),i=1,...,k-1$. Moreover, since any
$\zeta_{i}$ depends on $v_{i+1}$, the differential invariant $x,\zeta
_{0},...,\zeta_{k-2}$ are all functionally independent whereas $\zeta
_{k-1}=\zeta_{k-1}\left(  x,\zeta_{0},...,\zeta_{k-2}\right)  $. Therefore, in
terms of $\left\{  x,\zeta_{0},...,\zeta_{k-1}\right\}  $, $\Delta
(t,v,...,v_{k})=0$ can be rewritten in the form $\Delta_{red.}(x,\zeta
_{0},...,\zeta_{k-1})=0$. Now, since in $\mathcal{Y}^{\prime}$ the second
equation is integrable by the quadrature $w=\int\lambda dt$, the general
solution of $\mathcal{Y}^{\prime}$ can be obtained by solving first
$\Delta_{red.}(x,\zeta_{0},...,\zeta_{k-1})=0$ and then the auxiliary
$1$-order ODE $\zeta_{0}=\zeta_{0}(t,v,v_{1})$.

This proves the following

\begin{proposition}
The reduction of $\mathcal{Y}=\left\{  \Delta=0\right\}  $ via the $\lambda
$-symmetry $X^{[\lambda,k]}$ is conditioned to that of $\mathcal{Y}^{\prime
}=\left\{  \Delta=0,w_{1}=\lambda\right\}  $ via the nonlocal symmetry $Y$,
and vice-versa.
\end{proposition}

\begin{remark}
Notice that, since the reduction can be achieved just by means of the
invariants of $X^{[\lambda,k]}$, above discussion shows that the application
of MDI to $\mathcal{Y}=\left\{  \Delta=0\right\}  $ is independent from the
explicit determination of a solution to the reconstruction problem (see Remark
\ref{Remark_1}) for $X^{[\lambda,k]}$.
\end{remark}

A number of completely worked-out examples of reduction of ODEs by means of
the method of Proposition \ref{Rid-Muriel-Romero} are given in
\cite{Muriel-Romero-01} and \cite{Muriel-Romero-03}. Therefore the reader is
referred to these papers for a detailed list of examples. However, we conclude
this section with a couple of examples which provide an application of the
above nonlocal symmetry-reduction method.\newline\textbf{Example 4. }Let
$\mathcal{Y}$ be defined as%
\begin{equation}
v_{2}=-\frac{t^{2}}{4v^{3}}-v-\frac{1}{2v}.\label{Esempio-mu-Ro}%
\end{equation}
As shown in \cite{Muriel-Romero-01}, equation (\ref{Esempio-mu-Ro}) does not
have Lie point symmetries but, for $\lambda=t/v^{2}$ and $X=v\partial_{v}$,
admits the $\lambda$-symmetry $X^{[\lambda,2]}$. This $\lambda$-symmetry can
be recovered by a symmetry $Y$ of the system $\mathcal{Y}^{\prime}$, defined
by (\ref{Esempio-mu-Ro}) and $w_{1}=t/v^{2}$. In fact, by using
(\ref{Y_corresp})-(\ref{Y_corresp_bis}) and considering the solution
$\chi=2e^{w}$ of (\ref{eq_*}), one finds that $Y$ is the prolongation to
$J^{2}(\widetilde{\pi})$ of the vector field $e^{w}\left(  v\partial
_{v}-2\partial_{w}\right)  $. Now, the invariants $x$ and $\zeta_{0}$ reads
\[
x=t,\qquad\zeta_{0}=-\frac{v_{1}}{v}-\frac{t}{2v^{2}}%
\]
and, in terms of the system $\left\{  x,\zeta_{0},\zeta_{1}=d\zeta
_{0}/dx\right\}  $, (\ref{Esempio-mu-Ro}) reads
\begin{equation}
\zeta_{1}=\zeta_{0}^{2}+1.\label{mu-ro-int}%
\end{equation}
Therefore, since the general solution of (\ref{mu-ro-int}) is%
\[
\zeta_{0}=\tan(x+c_{1}),\qquad c_{1}\in\mathbb{R}%
\]
one can find the general solution of (\ref{Esempio-mu-Ro}) by solving the last
equation
\[
\frac{v_{t}}{v}+\frac{t}{2v^{2}}=-\tan(x+c_{1}).
\]
Since this equation can be linearized by the transformation $v\longmapsto
v^{2}$, the general solution to (\ref{Esempio-mu-Ro}) is%
\[
v=\pm\cos(t+c_{1})\sqrt{-\ln(\cos(t+c_{1}))-t\tan(t+c_{1})+c_{2}},\qquad
c_{1},c_{2}\in\mathbb{R}.
\]
\textbf{Example 5.} Let $\mathcal{Y}$ be the ODE (\ref{Esempio 3}) of Example
3. As observed above, $\mathcal{Y}$ admits a $\lambda$-symmetry which can be
recovered by the following (higher) symmetry of $\mathcal{Y}^{\prime}$ (the
system defined by (\ref{Esempio 3}) and $w_{1}=v+t/v$)%
\[
Y=\sum_{s}\left[  \widetilde{D}^{s}\left(  e^{w}v\right)  \partial_{v_{s}%
}+\widetilde{D}^{s}\left(  e^{w}\chi\right)  \partial_{w_{s}}\right]
\]
where $\chi=\chi(t,v,v_{1})$ is a solution of (\ref{pippone}). Now, the
invariants $x$ and $\zeta_{0}$ are just the invariants of $\overline{Y}$ (the
restriction of $Y$ to $\left(  \mathcal{Y}^{\prime}\right)  ^{\infty}$) which
depend only on $t,v,v_{1}$. Therefore, since $\overline{Y}$ simply reads as%
\[
\overline{Y}=e^{w}\left[  v\partial_{v}+\chi\partial_{w}+\left(  v_{1}%
+v^{2}+t\right)  \partial_{v_{1}}\right]  ,
\]
one can readily check that these invariants are%
\[
x=t,\qquad\zeta_{0}=\frac{v_{1}+t}{v}-v
\]
and, in terms of the system $\{x,\zeta_{0},\zeta_{1}=d\zeta_{0}/dx\}$,
(\ref{Esempio 3}) becomes $\zeta_{1}=0$.\newline It follows that $\zeta_{0}%
=c$, $c\in\mathbb{R}$, and then the solutions of (\ref{Esempio 3}) are
described by the generalized Riccati equation (see \cite{Ince})%
\[
v_{1}=v^{2}+cv-t.
\]

\section{Summary and concluding remarks}

The relevance of $\lambda$-symmetries, introduced first in
\cite{Muriel-Romero-01}, is due to the fact that a number of equations not
possessing Lie point symmetries can be reduced by a method which makes use of
$\lambda$-symmetries. Unfortunately, despite their name, $\lambda$-symmetries
of an ODE $\mathcal{Y}=\left\{  v_{k}=f\right\}  $ are not at all symmetries
(unless $\lambda=0$) and the $\lambda$-symmetry reduction method may appear to
be somehow unrelated to standard symmetry reduction methods.

We have shown it is not the case: according to the main result of this paper
(see Proposition \ref{Main_proposition}), $\lambda$-symmetries of
$\mathcal{Y}$ correspond to a special kind of nonlocal symmetries of
$\mathcal{Y}$.

In fact, following the approach to nonlocal symmetries based on the notion of
coverings \cite{Vin-et-al,Vino-89}, $\lambda$-symmetries can be obtained by
first embedding $\mathcal{Y}$ into the system $\mathcal{Y}^{\prime}=\left\{
v_{k}=f,w_{1}=\lambda\right\}  $ and then computing local symmetries $Y$ of
$\mathcal{Y}^{\prime}$ which have the form (\ref{pippo})-(\ref{pippo_2}) and
are such that $\left[  \partial_{w},Y\right]  =Y$.

As a consequence, one can show that (see section \ref{riduzione} above) the
$\lambda$-symmetry reduction method of Muriel and Romero readily follows from
the standard method of reduction by differential invariants applied to the
system $\mathcal{Y}^{\prime}$.

Finally, some remarks are in order. Since their first appearance, $\lambda
$-symmetries have stimulated new research on the reduction problem for
differential equations
\cite{Cicogna-Gaeta-Morando,Gaeta-Morando-2,Muriel-Olver,Pucci-Saccomandi}. In
particular, a generalization of $\lambda$-symmetries has been proposed in
\cite{Pucci-Saccomandi} through the so called telescopic vector fields.
However, as in the case of $\lambda$-symmetries, also this generalization can
be recovered by a nonlocal theory of symmetries. In fact, any telescopic
vector field can be recovered as a nonlocal symmetry of the form
(\ref{pippo})-(\ref{pippo_2}) such that $\xi=\xi(t,v,v_{1},w),\eta_{0}%
^{1}=\eta_{0}^{1}(t,v,v_{1},w)$ and $\left[  \partial_{w},Y\right]  =Y$.

\ack
The author would like to thank Giuseppe Gaeta and Paola Morando for their
comments and suggestions.



\section*{References}
\begin{thebibliography}{10}


\bibitem {AbrahamShrauner-Govinder-Leach}B. Abraham-Shrauner, K. S. Govinder
and P.G.L. Leach, \emph{Integration of second order ordinary differential
equations not possessing Lie point symmetries}, Phys. Lett. A \textbf{203},
169-174, 1995

\bibitem {Arnold}V. I. Arnold, \emph{Geometrical methods in the theory of
ordinary differential equations}, Springer, 1988

\bibitem {AVL}D.V. Alexseevskij, A.M. Vinogradov, V.V. Lychagin, \emph{Basic
ideas and concepts of differential geometry}, \textit{Geometry I --
Encyclopaedia of Mathematical Sciences vol. 28} (R.V. Gamkrelidze ed.),
Springer, 1991

\bibitem {Backlund}A. V. B\"{a}cklund, \emph{Ueber Flachentransformationen},
Math. Ann., \textbf{9}, 297-320, 1876

\bibitem {Bluman-Anco}G. W. Bluman and S. C. Anco, \emph{Symmetry and
Integration Methods for Differential Equations}, Springer, 2002

\bibitem {Bluman-Kumei}G. W. Bluman and S. Kumei, \emph{Symmetries and
differential equations}, Springer, 1989

\bibitem {Bluman-Reid}G. W. Bluman and G. J. Reid, \emph{New symmetries for
ordinary differential equations}, IMA J. Appl. Math., \textbf{40}, 87-94, 1988

\bibitem {Cicogna-Gaeta-Morando}G. Cicogna, G. Gaeta and P. Morando, \emph{On
the relation between standard and }$\mu$\emph{-symmetries for PDEs}, J. Phys.
A, \textbf{37}, 9467-9486, 2004

\bibitem {Gaeta-Morando-2}G. Gaeta and P. Morando, \emph{On the geometry of
}$\lambda$\emph{-symmetries and PDE reduction}, J. Phys. A, \textbf{37},
6955-6975, 2004

\bibitem {Gonzalez-Gonzalez}F. Gonz\'{a}lez-Gasc\'{o}n and A.
Gonz\'{a}lez-L\'{o}pez, \emph{Newtonian systems of differential equations,
integrable via quadratures, with trivial group of point symmetries}, Phys.
Lett. A, \textbf{129}, 153-156, 1988

\bibitem {Gonzalez}A. Gonz\'{a}lez-L\'{o}pez, \emph{Symmetry and integrability
by quadratures of ordinary differential equations}, Phys. Lett. A,
\textbf{45}, 190-194, 1988

\bibitem {Govinder-Leach}K. S. Govinder and P. G. L. Leach, \emph{A
group-theoretic approach to a class of second-order ordinary differential
equations not possessing Lie point symmetries}, J. Phys. A, \textbf{30},
2055-2068, 1997

\bibitem {Ibragimov-russo}N. H. Ibragimov, \emph{Essays in the Group Analysis
of Ordinary Differential Equations}, Matematika-Kibernetika, Znanie Publ.
Moscow, 1991 (Russian)

\bibitem {Ibragimov-trad}N. H. Ibragimov, \emph{Lie Group Analysis, Classical
Heritage}, ALGA Publications, 2004

\bibitem {Ince}E. L. Ince, \emph{Ordinary differential equations}, Dover
Publications, 1944

\bibitem {Konopelchenko}B. G. Konopelchenko and V. G. Mokhnachev, \emph{On the
group theoretical analysis of differential equations}, J. Phys. A,
\textbf{13}, 3113-3124, 1980

\bibitem {Kapcov}O. V. Kapcov, \emph{Extension of the symmetry of evolution
equations}, Sov. Math. Dokl., \textbf{25}, 173-176, 1982

\bibitem {Kras-Vin}I. S. Krasil'shchik and A. M. Vinogradov, \emph{Nonlocal
Trends in the Geometry of Differential Equations: Symmetries, Conservation
Laws, and B\"{a}cklund Transformations}, Acta Appl. Math. \textbf{15},
161-209, 1989

\bibitem {Muriel-Romero-01}C. Muriel and J. L. Romero, \emph{New methods of
reduction for ordinary differential equations}, IMA J. Appl. Math.,
\textbf{66}, 111-125, 2001

\bibitem {Muriel-Romero-03}C. Muriel and J. L. Romero, $C^{\infty}%
$\emph{-Symmetries and Reduction of Equations Without Lie Point Symmetries},
J. Lie Theory, \textbf{13}, 167-188, 2003

\bibitem {Muriel-Olver}C. Muriel, J. L. Romero and P. J. Olver,
\emph{Variational }$C^{\infty}$\emph{-symmetries and Euler-Lagrangian
equations}, J. Differential Equations \textbf{222}, 164-184, 2006

\bibitem {Olver1}P. J. Olver, \emph{Applications of lie groups to differential
equations}, Springer, 1993

\bibitem {Olver2}P. J. Olver, \emph{Equivalence, Invariance and Symmetry},
Cambridge University Press, 1995

\bibitem {Painleve}P. Painlev\'{e}, \emph{M\'{e}moire sur les \'{e}quations
diff\'{e}rentielles dont l'int\'{e}grale g\'{e}n\'{e}rale est uniforme}, Bull.
Soc. Math. France \textbf{28}, 201, 1900

\bibitem {Pucci-Saccomandi}E. Pucci and G. Saccomandi, \emph{On the reduction
methods for ordinary differential equations}, J. Phys. A, \textbf{35},
6145-6155, 2002

\bibitem {Smirnov}V. I. Smirnov, \emph{A course of higher mathematics},
Pergamon Press, 1964

\bibitem {Vin-et-al}A. M. Vinogradov et al., \emph{Symmetries and conservation
laws for differential equations of mathematical physics}, American
Mathematical Society, 1999

\bibitem {Vino-89}A. M. Vinogradov, \emph{Symmetries and Conservation Laws of
Partial Differential Equations: Basic Notions and Results}, Acta Appl. Math.
\textbf{15}, 3-21, 1989
\end{thebibliography}
\end{document}